\documentclass[conference]{IEEEtran}
\IEEEoverridecommandlockouts

\usepackage{amsmath,amssymb,amsfonts}
\usepackage{algorithmic}
\usepackage{graphicx}
\usepackage{color,listings}
\usepackage{enumitem}

\def\BibTeX{{\rm B\kern-.05em{\sc i\kern-.025em b}\kern-.08em
    T\kern-.1667em\lower.7ex\hbox{E}\kern-.125emX}}

\definecolor{asblue}{rgb}{0,0.2,0.6}
\definecolor{asred}{rgb}{0.8,0,0}
\definecolor{asgreen}{rgb}{0,0.4,0}
\definecolor{asorange}{rgb}{1.0,0.5,0.3}

\setlist[itemize]{noitemsep, topsep=2pt}

\newcommand{\Def}{\stackrel{\text{\tiny def}}{=}}

\NewDocumentCommand{\Half}{O{1}}{\frac{#1}{2}}

\newcommand{\Funct}[2]{#1\!\left(#2\right)}
\newcommand{\Funcf}[3]{#1\!\left(\frac{#2}{#3}\right)}

\newcommand{\Floor}[1]{\left\lfloor #1 \right\rfloor}
\newcommand{\Ceil}[1]{\left\lceil #1 \right\rceil}

\begin{document}

\title{Bitstream Organization for Parallel Entropy Coding on Neural Network-based Video Codecs\thanks{$^{*}$Qualcomm AI Research is an initiative of Qualcomm Technologies, Inc.}}

\author{\IEEEauthorblockN{
Amir Said}
\IEEEauthorblockA{\textit{Qualcomm AI Research$^{*}$} \\
San Diego, CA, USA \\
asaid@qti.qualcomm.com}
\and
\IEEEauthorblockN{
Hoang Le}
\IEEEauthorblockA{\textit{Qualcomm AI Research$^{*}$} \\
San Diego, CA, USA \\
hoanle@qti.qualcomm.com}
\and
\IEEEauthorblockN{
Farzad Farhadzadeh}
\IEEEauthorblockA{\textit{Qualcomm AI Research$^{*}$} \\
San Diego, CA, USA \\
ffarhadz@qti.qualcomm.com}
}

\maketitle

\begin{abstract}
Video compression systems must support increasing bandwidth and data throughput at low cost and power, and can be limited by entropy coding bottlenecks. Efficiency can be greatly improved by parallelizing coding, which can be done at much larger scales with new neural-based codecs, but with some compression loss related to data organization. We analyze the bit rate overhead needed to support multiple bitstreams for concurrent decoding, and for its minimization propose a method for compressing parallel-decoding entry points, using bidirectional bitstream packing, and a new form of jointly optimizing arithmetic coding termination. It is shown that those techniques significantly lower the overhead, making it easier to reduce it to a small fraction of the average bitstream size, like, for example, less than 1\% and 0.1\% when the average number of bitstream bytes is respectively larger than 95 and 1,200~bytes.
\end{abstract}


\section{Introduction}

Developers of video compression must support many new applications and quality requirements. For instance, there is increasing demand for streaming videos at higher resolutions, frame rates and dynamic range.
Those applications require high data throughput between and within devices, but since they are used in consumer products and mobile devices, they also need to minimize equipment cost, bandwidth, and power usage.

Multimedia compression~\cite{Woods:11:msp,Wiegand:11:src,Pearlman:11:dsc,Sze:14:hev,Wien:15:HEV} requires significant computational complexity, and the only way now to reduce costs and power is to parallelize computations. This is a consequence of physical limits constraining hardware design~\cite{Hennessy:19:nga}, and since those are irreversible, compression techniques must be updated or redesigned to allow more efficient massively parallel computations.

Some compression components can leverage concurrent signal processing operations for higher efficiency, but entropy coding is especially difficult to parallelize~\cite{Asanovic:06:lpc,Asanovic:08:vpc}, increasingly creating performance bottlenecks. There has been few works on clearly defining the problem and proposing new solutions.

While conventional video compression standards started to provide more features to enable parallel acceleration~\cite{Zhao:09:ped,Schwarz:14:bsp}, their compression methods are strongly based on sequentially exploiting data dependencies, and thus \emph{``higher speedups can be achieved [only] at the cost of decreased coding efficiency''}~\cite{George:22:emt}, which strongly limits parallelization.

More recently developed end-to-end neural video codecs~\cite{Balle:18:vic,Agustsson:20:ssf,Le:22:mob,Li:22:hst,Li:23:dcv,Rozendaal:24:mob}, on the other hand, use very different compression techniques, where deep-learning is used for designing systems with \emph{factorized priors,} meaning that data to be coded are statistically independent, and thus can in theory be entropy coded independently without loss.

\begin{figure}
\centering
\includegraphics[scale=0.6]{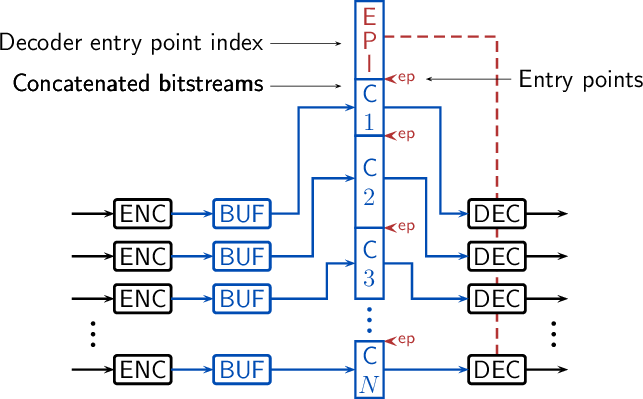}
\caption{\label{fg:MultiStream}Parallel coding using concatenated independent bitstreams for neural video codecs, with leading decoder entry point index.}
\end{figure}

However, this only corresponds to the theoretical data entropy. In practice, compressed data must be saved in bitstreams, and in general data elements cannot be randomly accessed for parallel decoding. A simple solution is to split the data to be encoded, and concatenate the resulting variable-length bitstreams, as illustrated in Figure~\ref{fg:MultiStream}. 

This requires attaching an \emph{entry point index} with pointers to the start of each independently coded bitstream.
This is a form of bit-rate overhead because the bytes used in this header should be added to compressed data size. A second overhead is with bits ``wasted'' to finish bitstreams to an integer number of bytes

\subsection{Paper contributions}

We show that, with a number $D$ of compressed data bytes, and $N_s<D$ concurrent execution threads, the relative overhead size compared to $D$ is in the form: 
\begin{equation}
 W(D, N_s; \alpha, \beta) \approx \frac{N_s [\alpha \log_2(D / N_s) + \beta]}{D}, \label{eq:OHForm}
\end{equation}
where $\alpha$ and $\beta$ are constants that we show to depend on data organization and index compression. For example, if we simply use $i$~bits for each index element, and have an average of $t$~bits for bitstream termination, we have $\alpha=0$ and $\beta=i+t$.

The objective of this paper is to propose and analyze new schemes to organize data and compress entry points that minimize the relative overhead, or equivalently, minimize factors $\alpha$ and $\beta$, enabling higher parallelization without increasing compression loss.

\begin{figure}
\includegraphics[scale=0.65]{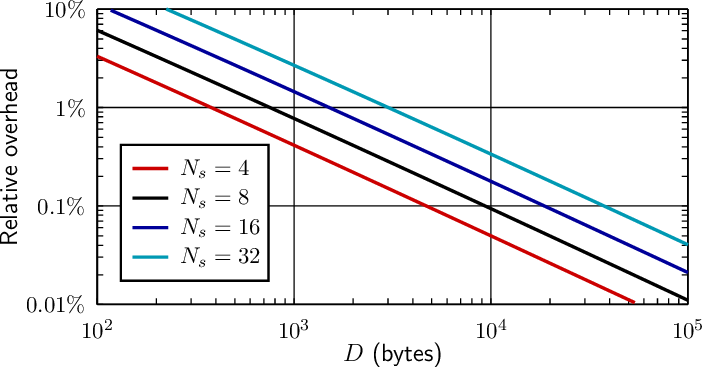}
\caption{\label{fg:MultStreamG}Graph showing an example of how the resulting data overhead depends on the data size $D$ and number of bitstreams $N_s$.}
\end{figure}

Figure~\ref{fg:MultStreamG} shows examples of plots of $W(D, N_s; \alpha, \beta)$ with values of $\alpha$ and $\beta$ corresponding to another combination of methods (actual parameters and full discussion in Section~\ref{sc:CombRes}). We can observe that since $W(D, N_s; \alpha, \beta)$ grows with $N_s$, we cannot increase parallelization without affecting compression, but we can use the graphs to choose a value of $N_s$ such that the relative overhead is acceptably small, which in turn depends on $\alpha$ and $\beta$.

The paper is organized as follows. In Section~\ref{sc:BackGrnd} we present background material to better define the problem and solutions. The optimization techniques are described in Sections \ref{sc:DataOrg} to \ref{sc:JointACT}, on using bidirectional bitstream packing, efficiently compressing parallel-decoding entry points, and a new form of jointly optimizing arithmetic coding termination. Experimental results and conclusions are in Sections \ref{sc:ExpRes} and \ref{sc:End}.

\section{Background}\label{sc:BackGrnd}

\subsection{Context-based entropy coding}\label{sc:CtxEC}

One of the main problems in the design of multimedia compression is that media signals are far from stationary. Thus, the main technical challenge is not in the process for converting information into bits, which can be done using conventional source coding methods, but in developing statistical data models and effective parameter estimation.

\begin{figure}[t]
\centering
\includegraphics[scale=0.6]{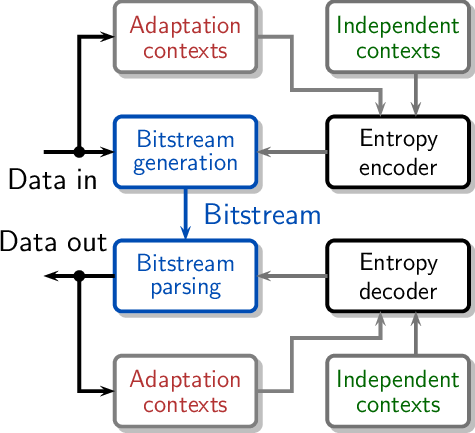}
\caption{\label{fg:EntCodec}General representation of context-based entropy coding.}
\end{figure}

\emph{Coding contexts} are used to represent different data models, and a large number of contexts and codes are commonly used. When considering parallel execution, we can differentiate between \emph{adaptation contexts}, that depend on the data being coded, and \emph{independent contexts} that do not, as shown in Figure~\ref{fg:EntCodec}. For generality, we also consider prediction as an adaptation context.
  
Each component in Figure~\ref{fg:EntCodec} can be modeled as a \emph{finite state machine} (FSM), and the whole encoder and decoder by a joint FSM~\cite{Said:15:cdo}. Bits added to a bitstream depend on both data and encoder state.

The decoder must duplicate the encoder's sequence of states to correctly parse the bitstream. For that reason, in general it is not possible to simply apply parallel decoders that are set to a pre-defined initial state, to arbitrary parts of the bitstream.

\subsection{Code self-synchronization}\label{sc:SelfSync}

Prefix codes (Golomb-Rice, Huffman, etc.)~\cite{Moffat:19:huf} have relatively simple FSMs, depending on single data inputs. For that reason they may be quickly \emph{self-synchronizing}, i.e., if decoding starts with an incorrect state, after outputting a few incorrect symbols the decoder \emph{probably} will recover and start decoding correctly. It has been shown that this property can be exploited for efficient parallel decoding, for example, of JPEG-compressed images~\cite{Klein:03:phd,Weissenberger:21:ajd}.

 However, it has not been applied to modern video codecs because they use more efficient methods, like arithmetic coding~\cite{Witten:87:acd,Said:03:acc,Said:04:iac} and ANS~\cite{Duda:15:tuo}, together with complex context selection, which when combined define FSMs with much larger complexity, exponentially larger state spaces, and thus have much lower probability of self-synchronization.

\subsection{Parallel coding with independent bitstreams}\label{sc:PCStreams}

The simplest form of parallel coding is shown in Figure~\ref{fg:MultiStream}: splitting data and generating independent bitstreams that can be concurrently decoded~\cite{Boliek:94:vhs,Marpe:10:eci,Said:15:cdo}. The start of each bitstreams is called \emph{entry point}, because decoders can begin parsing at those positions using pre-defined initial FSM states.
Note that, since final sizes of the bitstreams are not known \emph{a priori,} those bitstreams need to be buffered before their concatenation.

Alternatively, special \emph{start markers} can be added to the bitstream for identifying decoder entry points, but this requires a more complicated implementation to avoid generating erroneous markers.
Other data arrangements are possible when using synchronous vector instructions, but in this paper we assume more scalable asynchronous multithread execution.

This parallelization approach is straightforward, but constrained by compression degradation due to:
\begin{description}
 \item[\small (I)] For parallel decoding, adaptation contexts can only use data from the same bitstream, and compression cannot be improved by exploiting dependencies between bitstreams.
 \item[\small (II)] Bits are needed to store an index with the decoding entry point of each independent bitstream.
 \item[\small (III)] There are ``wasted'' bits in each bitstream, for example, to terminate arithmetic coding, and to generate an integer number of bytes.
\end{description}

\subsection{Data dependencies and neural-based codecs}\label{sc:NNCodecs}

Conventional video codecs exploit several types of data dependencies, like intra-frame prediction and entropy coding with adaptive contexts, that need to be disabled or limited for concurrent decoding. This causes fast compression degradation when the number of independent bitstreams increases.

\begin{figure}
\centering
\includegraphics[scale=0.6]{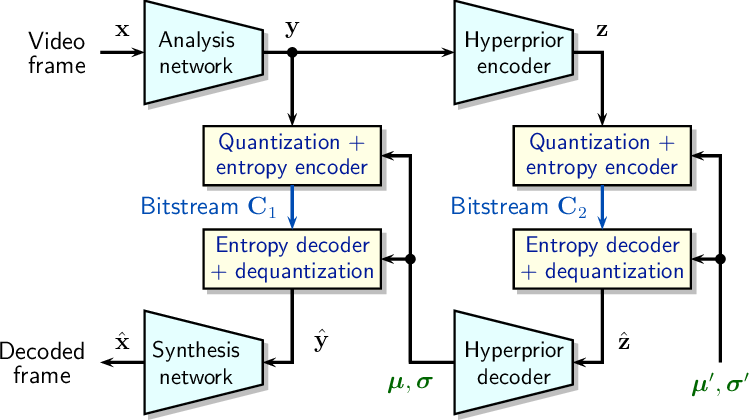}
\caption{\label{fg:NNHyper}Neural-based codec with hyperprior networks for computing coding parameters.}
\end{figure}

New video compression methods, based on deep-learning and neural networks, can overcome those limitations by changing the way those dependencies are exploited. Figure~\ref{fg:NNHyper} shows a simplified diagram of an end-to-end neural codec with \emph{hyperprior neural networks}~\cite{Balle:18:vic}, used in many video codecs~\cite{Agustsson:20:ssf,Le:22:mob,Li:22:hst,Li:23:dcv} (more information about neural codecs can be found in ref.~\cite{Ma:20:ivc}).

For our purposes, the important point is that in those codecs the tasks of prediction, bit rate allocation, and distribution estimation, which are commonly done sequentially by conventional codecs, are jointly performed by the hyperprior neural networks shown in the right side of Figure~\ref{fg:NNHyper}.

Those networks generate the parameters ${\boldsymbol\mu},{\boldsymbol\sigma}$, which are arrays of mean and standard deviation of normal random variables (factorized prior), and define all the parameters needed for coding the image or video data (represented by $\bf y$).

Note that those parameters, in turn, depend on side-information bit stream ${\bf C}_2$, which is computed from $\bf y$. What this means is that, to improve compression, neural codecs do exploit data dependencies, but they are computed in a very different manner, amenable to parallel computation and parallel entropy coding.

It is important to note that assumptions like using normal distributions for entropy coding are not observed \emph{a posteriori}, but are set as design objectives, and practically achieved via deep learning techniques~\cite{Kingma:14:aev,Goodfellow:16:dml}.

With this architecture, coding parameters ${\boldsymbol\mu},{\boldsymbol\sigma}$ define independent contexts (cf. Figure~\ref{fg:EntCodec}), and thus data in $\bf y$ can partitioned and coded in parallel without loss. This eliminates parallelization limitation~(I) in the list of Section~\ref{sc:PCStreams}.

In the next sections we propose methods to minimize factors (II) and (III), and in Section~\ref{sc:CombRes} analyze how their combined use affect overall compression.

\section{Data organization for parallel coding}\label{sc:DataOrg}

In this section we extend a technique that has been used for efficiently arranging two types of compressed data, to many parallel bitstreams as shown in Figure~\ref{fg:MultiStream}. It is used for halving the number of entry points, and thus significantly reduces their index size.

\subsection{Bidirectional bitstreams}\label{sc:ACBC}
\begin{figure}
\centering
\includegraphics[scale=0.56]{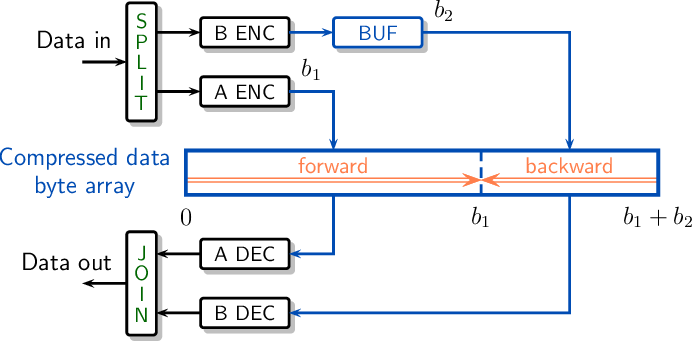}
\caption{\label{fg:BAStream}Forward and backward bitstream concatenation for combining arithmetic and binary coded data.}
\end{figure}

A technique for reducing coding complexity is to decompose data into components that need a more complex method like arithmetic coding, and the remaining data, which is saved directly in their binary representation~\cite{Marpe:03:cab,Said:05:eap}.

It has been observed that, with this scheme, it is not necessary to use extra bits for coding the number bytes used for binary coding if its data is saved in reverse order~\cite{Said:97:lcw,Valin:13:toc,Hallapuro:19:scc} and the two bitstreams are later concatenated, as shown in Figure~\ref{fg:BAStream}.

The decoders can pre-load two different types of data at the end of each bitstream, but with appropriate termination (cf. Section~\ref{sc:JointACT}), there are no decoding errors. This is possible because in multimedia compression the number of data symbols to be decoded is known, and the extra data is never parsed~\cite[\S4.1]{Valin:12:oac}. It is also easy to avoid invalid memory access by using sufficiently large buffers.

\subsection{Multiple forward and backward bitstreams}

\begin{figure}[t]
\centering
\includegraphics[width=84mm]{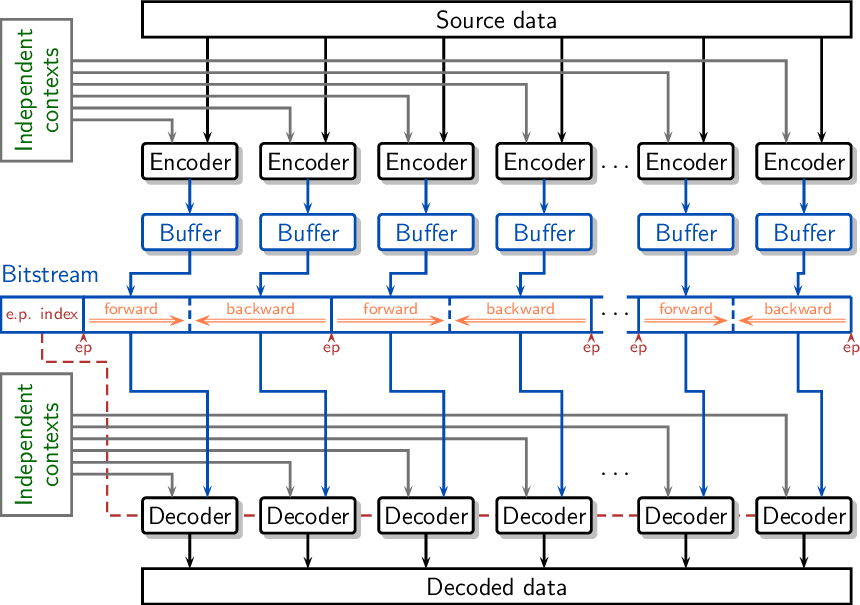}
\caption{\label{fg:FBBitstream}Architecture for parallel entropy coding using entry point index with multiple forward and backward bitstreams.}
\end{figure}

The compressed data arrangement described in the previous section can be extended to an arbitrary number of bitstream pairs, with the backward stream generated by arithmetic coding. The only changes in the arithmetic decoder implementations is that, after reading a byte, one increments and the other decrements a pointer, which is trivial to implement and does not increase coding complexity.

The combination of those techniques define the parallel entropy coding architecture shown in Figure~\ref{fg:FBBitstream}. This approach is more efficient when used with neural codecs because, as explained in Section~\ref{sc:NNCodecs}, they can use only need independent contexts.

In this scheme each entry point defines the position of the first byte in a forward bitstream, and the preceding byte is the first byte of a backward bitstream. Since each position is used for two bitstreams, the total number of entry points is halved. The first and last entry-point are exceptions, but in a compressed video file those positions are respectively at the end and beginning of frames indexes, so we can consider that the total number entry points is halved.

\section{Entry point index compression}\label{IdxComp}

Assuming $N_e$ parallel decoding entry points, as shown in Figure~\ref{fg:FBBitstream}, their index is defined by the number of bytes $b_i$ in each bitstream $C_i$, with entry point positions defined by the cumulative sums
\begin{equation}
 h_0 \equiv 0, \quad h_n \Def h_{n-1} + b_n = \sum_{i=1}^{n} b_i, \quad n = 1, 2, \ldots, N_e. \label{eq:EPArray}
\end{equation}

The average and minimum are defined as
\begin{equation}
 \bar{b} \Def \frac{1}{N_e} \sum_{i=1}^{N_e} b_i, \quad b_{\min} \Def \min\left( b_1, b_2, \ldots, b_{N_e} \right). \label{eq:MeanMin}
\end{equation}

To minimize the index overhead it is necessary to efficiently encode the sequences
\begin{equation}
 \mathcal{I} \Def \left( b_1, b_2, \ldots, b_{N_e} \right), \quad \mbox{or} \quad \mathcal{H} \Def \left( h_1, h_2, \ldots, h_{N_e} \right).
\end{equation}

Theoretically, this is a lossless data compression problem, requiring a statistical model and matched coding method. However, since the index size should be relatively very small, it may be preferable to avoid developing a specialized method for each case. 

One alternative, for example, is to simply save each element in $\mathcal{I}$ using the native precision (e.g. 32~bits). This is clearly inefficient, but commonly used because it is trivial to implement. In the next sections we show alternatives that are much more efficient \emph{and} very simple to implement.

\subsection{Universal codes}

Even without prior knowledge about $\mathcal{I}$, it is possible to have efficient compression by using ``universal''  methods, that combine coding with gathering information to improve compression, and that are capable of working well in a wide variety of cases.

Universal prefix codes proposed by Elias~\cite{Elias:75:ucs} are applicable to positive integers. The Elias $\gamma$ or “exp-Golomb” code, is popular and used in video compression standards AVC and HEVC. The number of bits it uses for coding a positive integer number $b$ is
\begin{equation}
  E_{\gamma} = 2 \Floor{log_2(b)} + 1 \; \mbox{ bits.}
\end{equation}

We can have better compression using a method developed to code all numbers in $\mathcal{I}$ together, like Binary Interpolative Compression (BIC)~\cite{Moffat:96:eci}, which can be applied directly to the sequence of entry point positions $\mathcal{H}$, and has an average number of bits equal to
\begin{equation}
  R_{\text{bic}} \approx \Funct{\log_2}{\bar{b}} + 2 \; \mbox{ bits/entry point}. \label{eq:BICrate}
\end{equation}

\subsection{Range-Tree Compression}\label{sc:RTC}

\begin{figure}
\centering
\includegraphics[width=85mm]{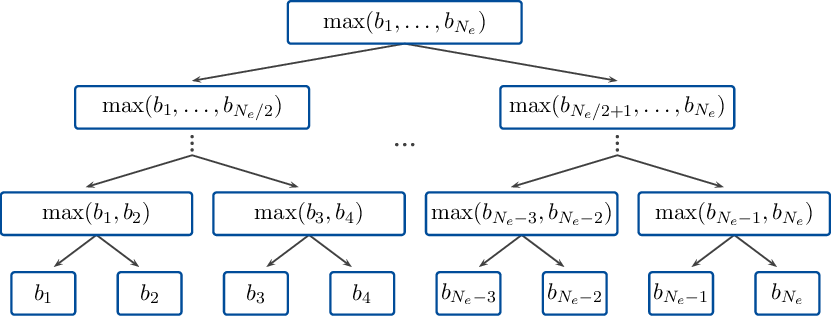}
\caption{\label{fg:RTC}Data used by Range-Tree Compression (RTC).}
\end{figure}

\begin{figure*}
\centering
\includegraphics[scale=0.75]{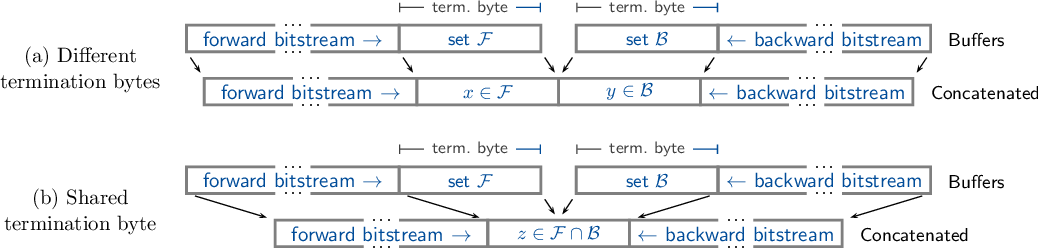}
\caption{\label{fg:FBMerge}Concatenations of forward and backward bitstreams with joint arithmetic coding termination.}
\end{figure*}

One limitation of BIC is that it is less efficient when values are tightly grouped around $\bar{b}$, and bit rate~(\ref{eq:BICrate}) becomes much larger than entropy. 
To cover all cases we propose using the tree-based coding approach of~\cite{Said:97:lcw} to design a simple universal coding method we call Range-Tree Compression (RTC).

Figure~\ref{fg:RTC} shows the data used by RTC, organized in a binary tree (for convenience we assume $N_e$ is a power of two). Managing tree data is simplified when information of a node is stored at position $i$ and information about its descendants is at positions $2i$ and $2i + 1$. 

Using this convention, we construct arrays with sizes $2N_e-1$ and $N_e$, containing maximum values and their selection, as
\begin{align}
  a_i & = \begin{cases}
    b_{i+1-N_e}, & N_e \leq i \leq 2 N_e - 1, \\
    \max(a_{2i}, a_{2i+1}), & 1 \leq i \leq N_e - 1,
  \end{cases} \\
  x_i & = \begin{cases}
    1, & a_{i} = a_{2i}, \\
    0, & a_{i} > a_{2i},
  \end{cases}, \quad i = 1, 2, \ldots, N_e.
\end{align}

Note that $\bf a$ is defined from other elements in the same array, but this simply means elements should be computed in reverse order.

With those definitions, we have the following properties
\begin{equation}
 a_{2i+1-x_i} = a_i, \quad b_{\min} \leq a_{2i+x_i} \leq a_i - x_i.
\end{equation}
which greatly simplify sequentially coding. When the value of $a_i$ is known, then we can
\begin{enumerate}
 \item Use only one bit to code $x_i$ and obtain $a_{2i+1-x_i}$.
 \item Code $a_{2i+x_i}$ using a simple method for compressing bounded integers values.
\end{enumerate}

A Python program implementing RTC with few lines of code is provided in the Appendix, and its compression is evaluated in Section~\ref{sc:IdxEval}.

\section{Joint arithmetic coding termination}\label{sc:JointACT}

As mentioned in Section~\ref{sc:PCStreams}, some extra bits are needed at the end of each bitstreams, and the total overhead increases with the number of bitstreams. In this section we show that, thanks to the bidirectional bitstream organization, arithmetic coding properties can be exploited to reduce the number of termination bits.

\subsection{Shared arithmetic coding termination bytes}

Recent video compression standards and neural-based codecs employ arithmetic coding (AC). Unlike earlier AC versions that process compressed data bits~\cite{Martin:79:rea,Witten:87:acd,Pennebaker:88:oqc}, more efficient modern implementations read and write blocks of several bits, commonly 8-bit bytes~\cite{Schindler:98:afr,Moffat:98:acr,Said:03:acc,Said:04:iac}.

When arithmetic encoding finishes, it needs to ``flush'' pending information and add bits to guarantee correct decoding. The next sections provide more information about this process. Here it suffices to consider that there is a set of \emph{valid termination bytes,} i.e., values that guarantee correct decoding.

Normally, a single encoder can arbitrarily choose any valid value. With bidirectional bitstreams, on the other hand, we have two encoders, and it is possible to make smarter joint decisions.

Calling $\mathcal{F}$ and $\mathcal{B}$ the sets of valid termination bytes for the forward and backward bitstreams, we have two options when doing concatenation, as shown in Figure~\ref{fg:FBMerge}. 
\begin{enumerate}
\item If $\mathcal{F} \cap \mathcal{B} = \emptyset$ then set termination bytes to any values $x \in \mathcal{F}$ and $y \in \mathcal{B}$, and concatenate bitstreams.
\item Otherwise, choose any value $z \in \mathcal{F} \cap \mathcal{B}$ for a termination byte that is shared by both bitstreams, saving 8~bits in the resulting concatenated bitstream.
\end{enumerate}

It is interesting to note that nonempty intersections are more probable when there are many values in $\mathcal{F}$ or $\mathcal{B}$, and those correspond to most ``wasted'' bits, i.e., this technique is most efficient on eliminating the worst cases.

In the next section we discuss how arithmetic coding bitstreams need to be terminated in general, and use those results in Section~\ref{sc:ACJoint} to define the sets $\mathcal{F}$ and $\mathcal{B}$ shown in Figure~\ref{fg:FBMerge}.

\subsection{Arithmetic coding termination}\label{sc:ACIntro}

Arithmetic coding principles and implementations can be found in several references~\cite{Martin:79:rea,Witten:87:acd,Pennebaker:88:oqc,Moffat:98:acr,Said:03:acc,Said:04:iac,Marpe:10:eci}. In this section we provide only the information required to clarify how sets of valid termination bytes values are defined.

The arithmetic encoder keeps a semi-closed interval (``range'' state) defining the fractional number of pending bits. In practice this interval is represented with integers, of precision depending on the implementation.

\begin{figure}
\centering
\includegraphics[scale=0.7]{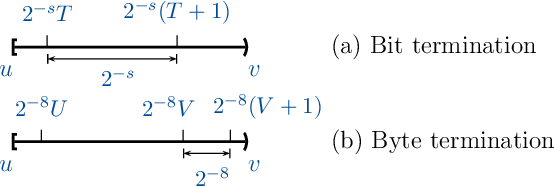}
\caption{\label{fg:ACTerm}Factors defining arithmetic coding termination.}
\end{figure}

We can describe the main principles, applicable to any implementation precision, by defining the encoder's semi-open final interval $[u, v)$ with real numbers, and normalized such that
\begin{equation}
 0 \leq u < 1, \quad u < v < 2, \quad 2^{-8} \leq v - u < 1. \label{eq:ACuvDef}
\end{equation}

In the scheme of Section~\ref{sc:ACBC}, the encoder must use termination bits such that, independently of the value of the following \emph{bits,} the corresponding decoder state is strictly within the final encoder interval, which guarantees correct decoding~\cite[\S5.1.5]{Valin:12:oac}.

As shown in Figure~\ref{fg:ACTerm}(a), this corresponds to finding an integer $T$ and the minimum integer $s$ that satisfy
\begin{equation}
 2^s u \leq T < T + 1 \leq 2^s v, \quad T, s \in \mathbb{N}.
\end{equation}
Note that at least one extra bit needs to be added since
\begin{equation}
 s \geq \Ceil{ - \log_2(v - u) } \geq 1. 
\end{equation}

When it is possible to choose all the bits in the termination byte, the objective is to instead guarantee correct decoding independently of the following \emph{bytes.} The corresponding condition is shown in Fig~\ref{fg:ACTerm}(b), where
\begin{equation}
  U \Def \Ceil{ 2^8 u } \quad V \Def \Floor{ 2^8 v } - 1. \label{eq:ACUVdef}
\end{equation}

The condition $V < U$ can occur, but only in some cases when $v - u < 2^{-7}$. In those cases an extra renormalization is needed, defining a new interval such that $U < V$. For this reason, in the following discussions we assume that $U \leq V$.

Integers $U$ and $V$ are sufficient to define correct termination, but do not necessarily correspond to final byte values because, following from~(\ref{eq:ACuvDef}), we can obtain values larger than 255. However, this simply corresponds to one \emph{addition carry}, which is a normal encoder operation. 

For convenience we use the modulo operation to define sets of byte values, with the implicit assumption that the encoder implements the required carry. The set of valid termination bytes is defined by
\begin{equation}
 \mathcal{V}(U, V) \Def \{ T \bmod 256: U \leq T \leq V, T \in \mathbb{N} \}. 
\end{equation}

\subsection{Set intersection and shared termination}\label{sc:ACJoint}

Using $U^{(F)}$, $V^{(F)}$, $U^{(B)}$, and $V^{(B)}$ to represent the limits of termination byte values defined in eq.~(\ref{eq:ACUVdef}), for respectively the forward and backward bitstreams, we have
\begin{equation}
 \mathcal{F} = \mathcal{V}\left( U^{(F)}, V^{(F)} \right), \quad
 \mathcal{B} = \mathcal{V}\left( U^{(B)}, V^{(B)} \right).
\end{equation}

\begin{figure}
\centering
\includegraphics[scale=0.7]{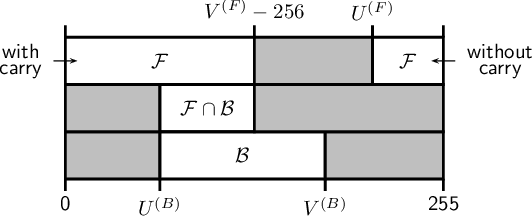}
\caption{\label{fg:ACTermEx}Example of sets of valid byte values for terminating arithmetic coding in forward and backward bitstreams.}
\end{figure}

Figure~\ref{fg:ACTermEx} provides an example, where the gray areas represent sets $\mathcal{F}$ and $\mathcal{B}$ and $\mathcal{F} \cap \mathcal{B}$, within the range of byte values [0, 255]. In this example the intersection is not empty, and the shared termination byte can be any integer $z \in [U^{(B)}, V^{(F)}-256]$, with a carry operation in the forward bitstream.

Note that this form of optimization uses only simple interval intersection determinations, which are computationally very simple, and only needs to be done once before concatenating forward and backward bitstreams. 

\subsection{Reverse bit order}

When individual bits are added to the binary bitstream discussed in Section~\ref{sc:ACBC}, the termination overhead will be smaller if, in the backward bitstream, both the bytes and bit are written in reverse order. 

For arithmetic coding the bit order is defined by the arithmetic operations, but we can still consider what happens when the bit order is reversed before reading and writing backward bitstream bytes (e.g., using table look-up).
Defining $\rho(n)$ as the function for bit reversal, we can use the same rules as before, but considering intersection of $\mathcal{F}$ with set
\begin{equation}
 \mathcal{B}_R = \{ \rho(n): n \in \mathcal{B} \}.
\end{equation}

\begin{figure}
\centering
\includegraphics[width=45mm]{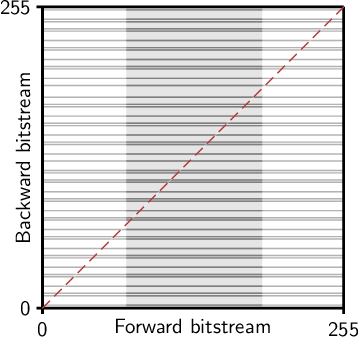}
\caption{\label{fg:ACDSets}Examples of valid joint termination byte values (gray areas), when the order of bits in the backward bitstream is reversed.}
\end{figure}

In this case set $\mathcal{B}_R$ does not contain only intervals, and appear as the multiple gray lines in the example shown in Figure~\ref{fg:ACDSets}, with the dashed red line in the center indicating the line where we can find intersection values that can be shared.

\section{Experimental results}\label{sc:ExpRes}

\subsection{Index compression evaluation}\label{sc:IdxEval}

When parallel coding is applied to neural video coding, there are wide variations in the distributions of bitstream sizes, depending on how data is split, quality settings, etc. One expected trend is to have values around a mode (distribution peak), with a longer tail of larger values.

\begin{figure}
\centering
\includegraphics[width=72mm]{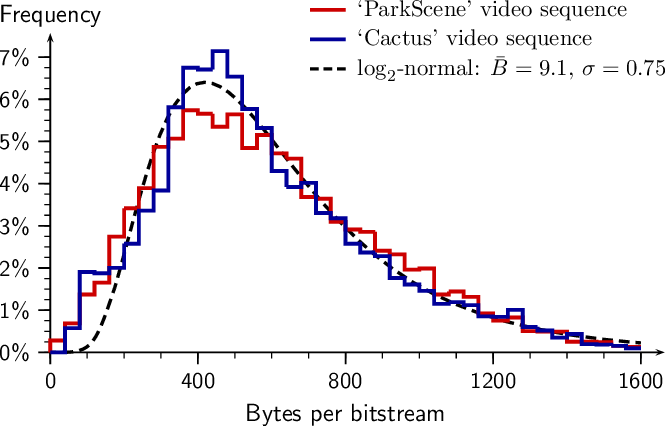}
\caption{\label{fg:ExpHist}Histograms of bitstream sizes from neural video compression on two video sequences, compared to a $\mbox{\bf log}_2$-normal distribution.}
\end{figure}

In Figure~\ref{fg:ExpHist} we have two examples from compressing 100 videos frames from VVC test sequences using the publicly available neural codec by Li \textit{et al.}~\cite{Li:22:hst} and 128~bitstreams/frame, where we can observe a remarkable similarity to a log-normal approximation.

Even though universal compression methods are not defined for specific distributions, we can get a good amount of insight by testing them using pseudo-random samples, and in our case from the log-normal probability distribution.

Since the numbers to be coded in $\mathcal{I}$ represent number of bits or bytes, we use a base-2 version of the log-normal distribution, defined from a normal random variable $Z \sim \mathcal{N}(\mu,\,\sigma^2)$, transformed to generate random variable $B = 2^Z$, with probability distribution function
\begin{equation}
 f_B(b; \mu, \sigma) = \frac{1}{b \ln(2) \sigma \sqrt{2 \pi}} \, \Funcf{\exp}{[\log_2(b) - \mu]^2}{2 \sigma^2}.
\end{equation}

Instead of using parameter $\mu$, we use the mean value
\begin{equation}
 \bar{B} = 2^{\mu + \ln(2) \sigma^2 / 2},
\end{equation}
since it corresponds to the expected value of $\bar{b}$, defined in~(\ref{eq:MeanMin}). With this notation the source entropy is
\begin{equation}
 H_B(\bar{B}, \sigma) = \Funct{\log_2}{\bar{B} \sigma \ln(2) \sqrt{2 e \pi}} - \frac{\ln(2)\sigma^2}{2}.
\end{equation}

\begin{table*}
\centering
\caption{\label{tb:IdxRes}Examples of results on neural video compression, 64~bitstreams for parallel decoding, 100~frames (ep = entry point).}
\begin{tabular}{|c||c||c|c|c|c||c|c|c|c||c|c|c|c|} \hline
\bf Bidir. & \bf Video seq.  & \multicolumn{4}{|c||}{\bf BQTerrace} & \multicolumn{4}{|c||}{\bf Cactus} & \multicolumn{4}{|c|}{\bf ParkScene}\\ \cline{2-14}
& \bf Quality & 0 & 1 & 2 & 3 & 0 & 1 & 2 & 3 & 0 & 1 & 2 & 3 \\ \hline \hline
\rule{0pt}{2.4ex} & $\bar{b}$ (bytes/ep)
 &  100.8 &  173.2 &  383.0 &  816.4 &   92.0 &  148.6 &  269.8 &  559.0 &  114.3 &  188.3 &  331.7 &  565.6 \\ \cline{2-14}
\rule{0pt}{2.4ex} & $\log_2(\bar{b})+2$
 &    8.7 &    9.4 &   10.6 &   11.7 &    8.5 &    9.2 &   10.1 &   11.1 &    8.8 &    9.6 &   10.4 &   11.1 \\
No & $R_{\text{rtc}}$ (bits/ep)
 &    8.1 &    9.0 &   10.3 &   11.4 &    8.0 &    8.8 &    9.8 &   10.8 &    8.3 &    9.2 &   10.1 &   10.8 \\
& $\hat{H}$ (bits/ep)
 &    7.9 &    8.8 &    9.8 &   10.3 &    7.8 &    8.6 &    9.3 &   10.0 &    8.2 &    9.0 &    9.6 &   10.1 \\ \cline{2-14}
& Overhead $W$ (\%)
 &   1.00 &   0.65 &   0.34 &   0.17 &   1.09 &   0.74 &   0.45 &   0.24 &   0.91 &   0.61 &   0.38 &   0.24 \\ \hline \hline
\rule{0pt}{2.4ex} & {\small $\bar{b}$} (bytes/ep)
 &  201.5 &  346.3 &  765.9 & 1632.7 &  183.9 &  297.1 &  539.5 & 1117.9 &  228.3 &  376.3 &  663.2 & 1131.0 \\ \cline{2-14}
\rule{0pt}{2.4ex} & $\log_2(\bar{b})+2$
 &    9.7 &   10.4 &   11.6 &   12.7 &    9.5 &   10.2 &   11.1 &   12.1 &    9.8 &   10.6 &   11.4 &   12.1 \\
Yes & $R_{\text{rtc}}$ (bits/ep)
 &    9.5 &   10.4 &   11.6 &   12.7 &    9.5 &   10.2 &   11.0 &   11.8 &    9.7 &   10.5 &   11.2 &   11.9 \\
& $\hat{H}$ (bits/ep)
 &    9.0 &    9.8 &   10.6 &   11.1 &    8.9 &    9.6 &   10.2 &   10.8 &    9.2 &    9.9 &   10.5 &   10.9 \\ \cline{2-14}
& Overhead $W$ (\%)
 &   0.59 &   0.38 &   0.19 &   0.10 &   0.65 &   0.43 &   0.25 &   0.13 &   0.53 &   0.35 &   0.21 &   0.13 \\ \hline
\end{tabular}
\end{table*}

\begin{figure}
\centering
\includegraphics[width=72mm]{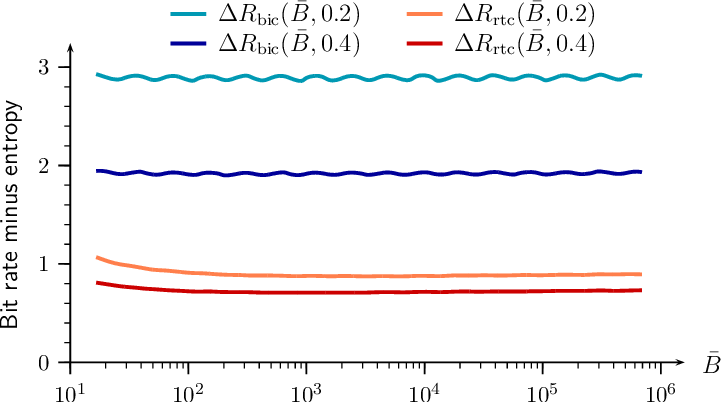}
\caption{\label{fg:RDMean}Redundancy of index compression methods applied to samples with $\mbox{\bf log}_2$-normal distribution, $\sigma = 0.2$ and $\sigma = 0.4$, and variable mean $\bar{B}$.}
\end{figure}

For our experiments on compressing samples from a $\log_2$-normal distribution with parameters $\bar{B}$ and $\sigma$, we use $R_{\text{bic}}$ and $R_{\text{rtc}}$ to represent bit rates obtained with BIC and RTC, and the difference between bit rate and source entropy (\emph{redundancy}) by
\begin{align}
 \Delta R_{\text{bic}}(\bar{B}, \sigma) & = R_{\text{bic}} - H_B(\bar{B}, \sigma), \\
 \Delta R_{\text{rtc}}(\bar{B}, \sigma) & = R_{\text{rtc}} - H_B(\bar{B}, \sigma). \nonumber
\end{align}

Compression methods like BIC and RTC are designed to work well independently of data magnitudes, and the graphs in Figure~\ref{fg:RDMean} show that indeed both BIC and RTC---despite not being specifically designed for log-normal distributions---give very consistent results when $\bar{B}$ varies over several orders of magnitude.

\begin{figure}
\centering
\includegraphics[width=72mm]{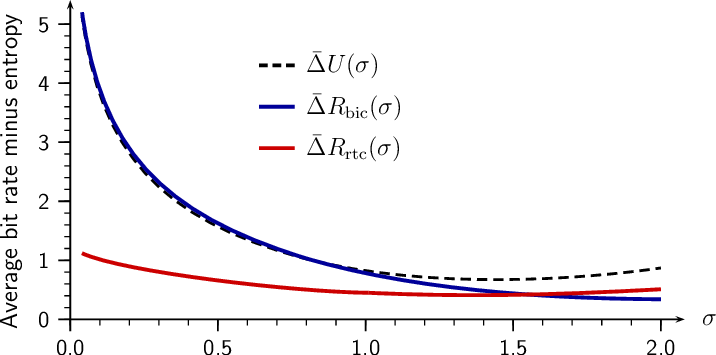}
\caption{\label{fg:RDStdDev}Average redundancies according to standard deviation $\sigma$.}
\end{figure}

On the other hand, we can also observe that there is are variations caused by $\sigma$, which is what RTC is meant to minimize. Defining the bit rate estimator from eq.~(\ref{eq:BICrate})
\begin{equation}
 \Delta U(\bar{B}, \sigma) = \Funct{\log_2}{\bar{B}} + 2 - H_B(\bar{B}, \sigma),
\end{equation}
and using the symbol $\bar{\Delta}$ to represent the average redundancy over the range $4 \leq \log_2(\bar{B}) \leq 20$, we measured its dependency on $\sigma$, and the results are shown in Figure~\ref{fg:RDStdDev}.

We can observe that eq.~(\ref{eq:BICrate}) is a good estimator of the BIC bit rates, and how the redundancy quickly grows when $\sigma$ decreases. The proposed RTC method, on the other hand, is more ``universal'' because its redundancy is nearly independent of the magnitudes \emph{and} variance. 

Table~\ref{tb:IdxRes} shows examples of results using the neural codec by Li \textit{et al.}~\cite{Li:22:hst}, on VVC test video sequences. For those experiments the tensor with latent variables (data to be compressed) of each frame is  ``flattened,'' and it its data is equally divided for encoding, to generate the $N_s$ independent bitstreams.

We can observe that~(\ref{eq:BICrate}) is again a very good estimator of the bit rates, obtained with RTC. In the table we also show $\hat{H}$, representing an estimate of the corresponding source entropy, and roughly $R_{\text{rtc}} < \hat{H}+1.5$.

\subsection{Joint termination evaluation}

For testing the AC termination described in Section~\ref{sc:JointACT} we used the implementation with 32-bit arithmetic available at the web site from~\cite{Pearlman:11:dsc} (resources/Software for students{$^\dagger$}).
\let\thefootnote\relax\footnote{
$^\dagger$https://www.cambridge.org/us/academic/subjects/engineering/communications-and-signal-processing/digital-signal-compression-principles-and-practice}

\begin{table}
\centering
\caption{\label{tb:ACTerm}Overhead from byte-based AC termination.}
\begin{tabular}{|c|c|c|c|c|} \hline
\bf Bidirectional  &\bf Reversed &\bf  Share &\bf Avrg. extra bits \\
\bf byte packing &\bf   bits   &\bf  ratio &\bf  per bitstream $\bar{T}$ \\ \hline \hline
     No     &     ---     &    ---    &        4.56         \\ \hline
     Yes    &     No      &    45\%   &        2.77         \\ \hline
     Yes    &     Yes     &    69\%   &        1.78         \\ \hline
\end{tabular}
\end{table}

Since these tests are only about the AC process, the results are normally independent of the data. Thus, AC termination was tested by coding pseudo-random binary samples with varying probabilities, and  results are shown in Table~\ref{tb:ACTerm}, where the \emph{share ratio} represents the fraction of terminations where a byte is shared, as in Figure~\ref{fg:FBMerge}(b).

Thanks to bidirectional byte packing and joint byte termination, and without the more complex reversal of bits in the backward bitstream, in 45\% of the cases the final byte can be shared, and the average number of extra bits per bitstream can be reduced from 4.56 to 2.77. With bit reversal the ratio increases to 69\% and the average decreases to 1.78~bits.

\subsection{Total parallelization overhead}\label{sc:CombRes}

We can combine the previous results to evaluate how they affect the total parallelization overhead, corresponding to eq.~(\ref{eq:OHForm}). We use the following acronyms to represent index formats, and to represent data organization that is combined with optimized AC termination
\begin{description}
\item[{\footnotesize \sf I32}] -- entry points represented with 32~bit values. 
\item[{\footnotesize \sf RTC}] -- entry points compressed with RTC. 
\item[{\footnotesize \sf UNI}] -- unidirectional byte packing.
\item[{\footnotesize \sf F+B}] -- bidirectional byte packing with same bit order within bytes. 
\item[{\footnotesize \sf F+R}] -- bidirectional byte packing with reversed bit order in the backward stream. 
\end{description}
 
Eq.~(\ref{eq:OHForm}) results from adding the following overhead terms
\begin{itemize}
 \item From experimental results in Figure~\ref{fg:RDStdDev} and Table~\ref{tb:IdxRes}: when using RTC the average overhead is commonly well-approximated by eq.~(\ref{eq:BICrate}), i.e., $\log_2(\bar{b})+2$ per entry point.
 \item From coding simulations we determined that the average AC termination overhead per bitstream is $\bar{T}$, shown in table~\ref{tb:ACTerm}.
\end{itemize}

With unidirectional byte packing the number of bitstreams is equal to the number of entry point, i.e., $N_s=N_e$. With bidirectional byte packing we have $N_s=2N_e$, and given the total number of bytes $D$, 
\begin{equation}
  \frac{N_e [\log_2(D/N_e) + 2] + N_s \bar{T}}{8D} = \frac{N_s [\log_2(D/N_s) + 3 + 2 \bar{T}]}{16D}. \nonumber
\end{equation}
and with this transformation, and substitution of factors with the obtained numerical values, we obtain the set of factor values $\alpha$ and $\beta$ in eq.~(\ref{eq:OHForm}) shown in Table~\ref{tb:AlphaBeta}. The lines shown in Figure~\ref{fg:MultStreamG} correspond to case {\sf F+R} \& {\sf RTC}.
 
\begin{figure}
\centering
\includegraphics[width=85mm]{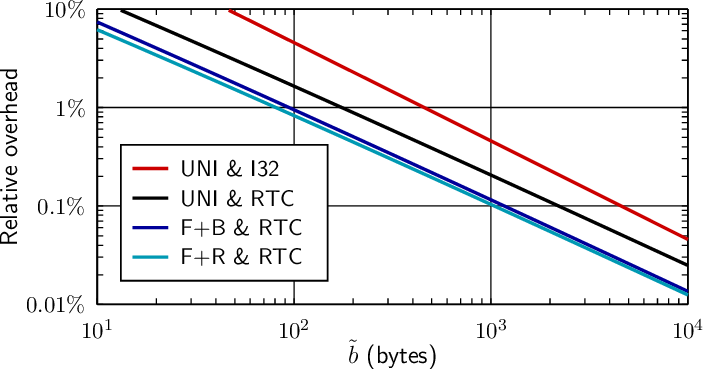}
\caption{\label{fg:RelOverhead}Relative overhead from factors in Table~\ref{tb:AlphaBeta}.}
\end{figure}

Observing that eq.~(\ref{eq:OHForm}) depends only on average bitstream size $\tilde{b} = D / N_s$, we can rewrite it as
\begin{equation}
 W(\tilde{b}; \alpha, \beta) \approx \frac{\alpha \log_2(\tilde{b}) + \beta}{\tilde{b}}, \label{eq:OHFormB}
\end{equation}
and plots of this function, using values in Table~\ref{tb:AlphaBeta}, are shown in Figure~\ref{fg:RelOverhead}.

\begin{table}
\centering
\caption{\label{tb:AlphaBeta}Overhead factors for different index coding and data organization.}
\begin{tabular}{|c|c|c|c|} \hline
\bf Order &\bf Index  & \boldmath $\alpha$ & \boldmath $\beta$ \\ \hline  \hline
\sf UNI   &\sf I32    &        0           &       4.57        \\ \hline
\sf UNI   &\sf RTC    &      1 / 8         &       0.82        \\ \hline
\sf F+B   &\sf RTC    &      1 / 16        &       0.53        \\ \hline
\sf F+R   &\sf RTC    &      1 / 16        &       0.41        \\ \hline
\end{tabular}
\end{table}

\section{Conclusions}\label{sc:End}

As shown in Figure~\ref{fg:RelOverhead}, our method significantly reduces the size of the overheads needed for parallel entropy coding. When the average number of bytes per bitstream is sufficiently large, the relative overhead can be small even with an inefficient index format, like {\sf I32}. In the same figure we can see that the proposed RTC index compression method significantly reduces the overhead.

Despite being extremely simple to implement, it is shown by experimental results in Figs. \ref{fg:RDMean}, \ref{fg:RDStdDev}, and Table~\ref{tb:IdxRes}, that RTC yields consistently good performance nearly independently of the scales and randomness (measured by entropy).

Employing bidirectional byte packing produces another significant reduction because:
\begin{itemize}
 \item It halves the number of elements in the entry point index, and as shown by results of Table~\ref{tb:IdxRes}, together with RTC consistently yields about 40\% index overhead reduction.
 \item It enables a more efficient form of jointly terminating two arithmetic coding bitstreams, which as shown in Table~\ref{tb:ACTerm}, reduce the termination overhead by 40\% with simple modifications and 60\% with an equally simple, but computationally more expensive AC modification.
\end{itemize}


\appendix
\subsection{RTC implementation}\label{sc:IdxProg}

\lstloadlanguages{Python}

This appendix presents a Python implementation of the Range-Tree Compression (RTC) method for coding the array with number of bytes in each bidirectional bitstream, which defines the index of decoder entry points.

The objective is to show that RTC is very simple, requiring only a few lines of Python code. At the same time, as shown by the experimental results, it is quite effective and produces consistently good compression results in a very wide range of situations and data distributions.

\lstset{
 language=Python,
 basicstyle=\footnotesize, 
 keywordstyle=\color{asblue}\bfseries,
 identifierstyle=, 
 commentstyle=\color{asred}, 
 stringstyle=\ttfamily, 
 frame=single,
 showstringspaces=false} 

We assume that functions \texttt{pack\_bit(x)} and \texttt{unpack\_bit()}, used for saving and retrieving individual bits to and from a bitstream are already implemented. They are quite simple, and details do not need to be repeated here.

Two other auxiliary functions, \texttt{pack(n, u)} and \texttt{unpack(u)}, are used for encoding integer $n$ in a given range defined by $u$, i.e.,
\begin{equation}
 n \in \{0, 1, 2, \ldots, u - 1\},
\end{equation}
assuming values are equally probable. This can be done with a prefix code~\cite{Moffat:19:huf} that assigns codewords using $\Floor{\log_2 u}$ or $\Ceil{\log_2 u}$ bits. There are many ways to do this, but one of the simplest, which does not require computing $\Floor{\log_2 u}$, is based on a bisection search and is shown below.

\begin{lstlisting}
def pack(n, u):
  a, b, m = 0, u, u // 2
  while a != m:
    if n < m: pack_bit(1);  b = m
    else:     pack_bit(0);  a = m
    m = (a + b) // 2  
\end{lstlisting}

\begin{lstlisting}
def unpack(u):
  a, b, m = 0, u, u // 2
  while a != m:
    if unpack_bit(): b = m
    else:            a = m
    m = (a + b) // 2  
  return m  
\end{lstlisting}

Below is the RTC encoder implementation, using the same notation of the paper. The parameters are the number of data elements to be coded \texttt{N}, the array with data \texttt{b}, and an upper bound on all values \texttt{T}. It is assumed \texttt{N} is a power of two.

\begin{lstlisting}
def RTC_encode(N, b, T):        # encoder function
  v, y = [0] * (N * 2), [0] * N # initialization
  v[N:2*N] = b[0:N]
  b_min = min(b)                # computation of b_min
  for i in range(N - 1, 0, -1): # and arrays v and y
    if v[2*i] >= v[2*i+1]: y[i], v[i] = 1, v[2*i]
    else:                  y[i], v[i] = 0, v[2*i+1]
  pack(v[1], T)                 # encoding v[1] and b_min
  pack(b_min, v[1])
  for i in range(1, N):         # progressive coding of
    if v[i] != b_min:           # tree-organized data
      pack_bit(y[i])
      pack(v[i] - v[2*i+y[i]] + y[i] - 1,\
           v[i] - b_min + y[i])
\end{lstlisting}

This function can be easily optimized, and when \texttt{N} is not a power of two, we can for instance pad the array with \texttt{b\_min}.

The corresponding decoder, shown below, does not need two arrays, since it reuses memory initially used for range-tree data to save the final decoded data.

\begin{lstlisting}
def RTC_decode(N, T):           # decoder function
  b = [0] * N                   # initialization
  b[1]  = unpack(T)             # decoding v[1] and b_min
  b_min = unpack(b[1])
  for i in range(1, N):         # in-place array decoding
    j = 2 * i if 2 * i < N else 2 * i - N
    b[j] = b[j+1] = b[i]
    if b[i] != b_min:
      y = unpack_bit()
      b[j+y] -= unpack(b[i] - b_min + y) - y + 1
  return b                      # return decoded array
\end{lstlisting}


\end{document}